\documentstyle[12pt]{article}
\def\l#1#2{\raisebox{.2ex}{$\displaystyle
  \mathop{#1}^{{\scriptstyle #2}\rightarrow}$}}
\def\r#1#2{\raisebox{.2ex}{$\displaystyle
 \mathop{#1}^{\leftarrow {\scriptstyle #2}}$}}

\makeatletter
\def\eqnarray{\stepcounter{equation}\let\@currentlabel=\theequation
\global\@eqnswtrue
\global\@eqcnt\z@\tabskip\@centering\let\\=\@eqncr
$$\halign to \displaywidth\bgroup\@eqnsel\hskip\@centering
  $\displaystyle\tabskip\z@{##}$&\global\@eqcnt\@ne
  \hfil$\displaystyle{\hbox{}##\hbox{}}$\hfil
  &\global\@eqcnt\tw@ $\displaystyle\tabskip\z@
  {##}$\hfil\tabskip\@centering&\llap{##}\tabskip\z@\cr}
\@addtoreset{equation}{section}
  \def\theequation{\thesection.\arabic{equation}}
\makeatother

\begin{document}

\renewcommand{\thefootnote}{\fnsymbol{footnote}}
\newpage
\setcounter{page}{0}
\pagestyle{empty}
\begin{flushright}
{January 1997}\\
{JINR E2-97-21}\\
{solv-int/9701019}
\end{flushright}
\vfill

\begin{center}
{\LARGE {\bf The solution of the $N=2$ supersymmetric}}\\[0.3cm]
{\LARGE {\bf f-Toda chain with fixed ends}}\\[1cm]

{\large A.N. Leznov$^{a,1}$ and A. Sorin$^{b,2}$}
{}~\\
\quad \\
{\em {~$~^{(a)}$ Institute for High Energy Physics,}}\\
{\em 142284 Protvino, Moscow Region, Russia}\\
{\em {~$~^{(b)}$ Bogoliubov Laboratory of Theoretical Physics, JINR,}}\\
{\em 141980 Dubna, Moscow Region, Russia}~\quad\\

\end{center}

\vfill

\centerline{ {\bf Abstract}}

The integrability of the recently introduced $N=2$ supersymmetric f-Toda
chain, under appropriate boundary conditions, is proven. The recurrent
formulae for its general solutions are derived. As an example, the
solution for the simplest case of boundary conditions is presented in
explicit form.

\vfill
{\em e-mail:\\
1) leznov@mx.ihep.su\\
2) sorin@thsun1.jinr.ru }
\newpage
\pagestyle{plain}
\renewcommand{\thefootnote}{\arabic{footnote}}
\setcounter{footnote}{0}

\section{Introduction}
Quite recently, we introduced the minimal $N=2$ supersymmetric
extension of the Toda chain called f-Toda \cite{dls}. We have shown that
the $N=2$ supersymmetric Nonlinear Schr\"odinger (NLS) hierarchy in the
$N=2$ superspace \cite{kst} possesses a discrete group of integrable
mappings \cite{fl} which is equivalent to the f-Toda chain.
Its origin was analyzed in \cite{s}.
In contrast to other chains, the f-Toda chain is not algebraically
solvable. Such type of chains has not been considered in the literature
before.

The goal of the present Letter is to prove the integrability of the f-Toda
chain and to construct its general solution under appropriate boundary
conditions, including the condition corresponding to fixed ends. This last
problem is equivalent to the problem of constructing
multi-soliton solutions for each integrable system belonging to the $N=2$
super-NLS hierarchy.

\section{The $N=2$ super-NLS and f-Toda superfield equations}

In this section, following \cite{dls}, we briefly describe the f-Toda
chain equations and their relation to the $N=2$ super-NLS hierarchy.

Let us proceed with the $N=2$ super-NLS equation \cite{kst}
\begin{equation}
\frac{\partial f}{\partial t}=f~''+2D(f\overline f~ \overline Df),\quad
\frac{\partial \overline f}{\partial t}=-\overline f~''+
2\overline D(f\overline f D \overline f),
\label{1}
\end{equation}
where $f(x,\theta,\overline \theta)$ and $\overline f(x,\theta,\overline
\theta)$ are chiral and antichiral fermionic superfields,
\begin{equation}
Df= 0, \quad \overline D~ \overline f=0,
\label{chir}
\end{equation}
respectively, in the $N=2$ superspace with one bosonic $z$ and two
fermionic $\theta,\overline \theta$ coordinates\footnote{The sign ${}~'$
means the derivative with respect to $z$.}; $D$ and $\overline D$ are the
$N=2$ supersymmetric fermionic covariant derivatives
\begin{equation}
D=\frac{\partial}{\partial \theta}-
\frac{1}{2}\overline \theta \frac{\partial}{\partial z},
\quad \overline D=\frac{\partial}{\partial \overline \theta}-
\frac{1}{2}\theta\frac{\partial} {\partial z}, \quad
D^2= {\overline D}^2=0, \quad \{D,\overline D\} = -
\frac{\partial}{\partial z}
\equiv - {\partial}.
\end{equation}
The chirality condition (\ref{chir}) reduces the number of
independent components of the fermionic
superfield $f$ ($\overline f$) from four to two.

Equations (\ref{1}), as well as each system of the
equations belonging to the $N=2$ super-NLS hierarchy, are invariant with
respect to the following discrete f-Toda transformation \cite{dls}:
\begin{equation}
\r {f}{} \r {\overline f}{}-f \overline f=
(\ln (\overline D \r {f}{} D \overline f))~'.
\label{2}
\end{equation}
This relation is the definition of mapping, i.e., the rule determining the
correspondence between two initial superfields $f$ and $\overline f$, and two
final ones, $\r {f}{}$ and $ \r {\overline f}{}$. The notation $\r {f}{}$
($\l {f}{}$) means that the index of the superfield $f$ is shifted by
$+1$ $(-1)$, i.e., $\r {f_k}{}\equiv f_{k+1}$ ($\l {f_k}{} \equiv
f_{k-1}$), and it denotes the action of the direct (inverse) f-Toda
transformation applied to the superfield $f$. The invariance of the $N=2$
super-NLS hierarchy with respect to the transformation (\ref{2}) can be
checked directly, and it means, in particular, that the right-hand sides
of eqs. (\ref{1}) are the solutions of the symmetry equation\footnote{Let
us recall that the symmetry equation for a given system can be obtained by
differentiation of the system with respect to an arbitrary parameter.}
corresponding to the mapping (\ref{2}) \cite{dls}.

\section{The component form and boundary conditions}
The $N=2$ superfield form (\ref{2}) of the f-Toda mapping is not
very suitable for actual calculations and we rewrite it in terms
of the superfield components defined as
\begin{eqnarray}
v_k = D {\overline f}_k|, \quad
{\overline {\psi}}_k =  D {\overline f}_k|, \quad
u_k  = {\overline D}f_k|, \quad
{\psi}_k = {\overline D} f_k|,
\label{com1}
\end{eqnarray}
where the index $k \in Z$ and $|$ means the
$({\theta}, {\overline\theta}) \rightarrow 0 $ limit.
In terms of these components, eq. (\ref{2}) becomes
the infinite-dimensional chain of equations
\begin{equation}
(\ln(u_{k+1}v_k))~'=\psi_{k+1}\overline \psi_{k+1}-\psi_k \overline \psi_k,
\label{b1}
\end{equation}
\begin{equation}
-({ \psi_{k+1}~'\over u_{k+1}})~'=-v_{k+1}\psi_{k+1}+v_k\psi_k,\quad
\label{3}
\end{equation}
\begin{equation}
-({\overline \psi_k~'\over v_k})~'=u_{k+1}\overline \psi_{k+1}-
u_k \overline \psi_k,
\label{33}
\end{equation}
\begin{equation}
-(\ln v_k)~''=u_{k+1}v_{k+1}-u_{k}v_{k}-
\psi_{k+1}~'~\overline \psi_{k+1}+\psi_k~'~\overline \psi_k.
\label{b2}
\end{equation}
Each point of this chain contains four functions: two bosonic
$u_k,v_k$ and two fermionic $\psi_k,\overline \psi_k$ functions. In the
bosonic limit (i.e., as $\psi_k \rightarrow 0$ and
$\overline \psi_k \rightarrow 0)$ it is equivalent to the
usual Toda chain\footnote{In the bosonic limit, eq. (\ref{b1})
admits the obvious solution $u_{k+1} \sim \frac{1}{v_k}$ which, being
substituted into eq. (\ref{b2}), produces the well-known representation for
the usual Toda chain equations.} \cite{t}. This is the reason why the chain
(\ref{2}) is called the fermionic Toda or f-Toda chain \cite{dls}.

For completeness, let us also present the component form of the global
$N=2$ supertranslations
\begin{eqnarray}
&& v_k \rightarrow v_k-
\varepsilon \overline \psi_k~' -
\frac{1}{2}\varepsilon \overline \varepsilon v_k~',\quad
\overline \psi_k \rightarrow \overline \psi_k+\overline \varepsilon v_k +
\frac{1}{2}\varepsilon \overline \varepsilon {\overline \psi}_k~' ,
\nonumber\\
&& u_k \rightarrow u_k-\overline \varepsilon \psi_k~' +
\frac{1}{2}\varepsilon \overline \varepsilon u_k~',\quad
\psi_k \rightarrow  \psi_k+ \varepsilon u_k -
\frac{1}{2}\varepsilon \overline \varepsilon {\psi}_k~'
\label{suptrans}
\end{eqnarray}
under which eqs. (\ref{b1})--(\ref{b2}) remain invariant. Here,
$\varepsilon$ and $\overline \varepsilon$ are constant fermionic parameters.
Equations (\ref{b1})--(\ref{b2}) also possess the
inner discrete automorphism $\sigma_m$ ($m \in Z$) with the properties
\begin{eqnarray}
\sigma_m u_k {\sigma}^{-1}_m= v_{m-k}, \quad
\sigma_m v_k {\sigma}^{-1}_m= u_{m-k} , \quad
\sigma_m \psi_k {\sigma}^{-1}_m=\overline \psi_{m-k}, \quad
\sigma_m {\overline \psi_k} {\sigma}^{-1}_m= \psi_{m-k},
\label{auto1}
\end{eqnarray}
which will be useful in what follows. Eqs. (\ref{auto1}) can also be
easily represented in terms of the superfields and superspace coordinates
\begin{eqnarray}
\sigma_m f_k {\sigma}^{-1}_m=\overline f_{m-k}, \quad
\sigma_m {\overline f_k} {\sigma}^{-1}_m= f_{m-k}, \quad
\sigma_m z {\sigma}^{-1}_m= z, \quad
\sigma_m \theta {\sigma}^{-1}_m= \overline \theta, \quad
\sigma_m \overline \theta {\sigma}^{-1}_m= \theta.
\label{autos1}
\end{eqnarray}

Let us discuss the boundary conditions which can supplement
eqs. (\ref{b1})--(\ref{b2}).

We call the system (\ref{b1})--(\ref{b2}) with additional manifest $N=2$
supersymmetric boundary conditions
\begin{equation}
f_0=0  \Leftrightarrow  u_0=\psi_0=0 , \quad k \geq 0 ,
\label{bc1}
\end{equation}
or
\begin{equation}
\overline f_M=0 \Leftrightarrow  v_M=\overline \psi_M=0, \quad  k \leq M,
\label{bc2}
\end{equation}
where $M \in Z_{+}$, the f-Toda chain interrupted from the
left or from the right, respectively. The case where these conditions are
satisfied simultaneously corresponds to the f-Toda chain with fixed ends.

Application of the transformation $\sigma_M$ (\ref{auto1}) to the f-Toda
chain interrupted from the left transforms it into the f-Toda chain
interrupted from the right, and vice versa. Due to their relation, it is
sufficient to consider only one of them; for concreteness, we discuss the
f-Toda chain interrupted from the left in what follows. As concerns the
f-Toda chain with fixed ends, it is invariant with respect to $\sigma_M$,
or, in other words, it possesses the inner automorphism $\sigma_M$.

\section{The integrability of the f-Toda chain interrupted from the left}
In this section, we construct the recurrent procedure for solving
of the f-Toda chain equations (\ref{b1})--(\ref{b2}) with the boundary
conditions (\ref{bc1}).

\subsection{The conservation laws}
Our first goal is to resolve the f-Toda chain interrupted from the left.
In other words, we assume that the boundary condition (\ref{bc1}) is
satisfied and $v_0,\overline \psi_0$ are arbitrary functions. The problem
consists in expressing, using the f-Toda chain equations, all of the
functions $u_k$, $v_k$, $\psi_k$ and $\overline \psi_k$ at $ k \geq 1$ in
terms of the functions $v_0$ and $\overline \psi_0$, as well as a number
of additional constants which can arise in integrating the equations.

Let us start with a discussion of the conservation laws which are
relevant to the problem under consideration.

By a direct check, one can verify that the following set of conservation
laws:
\begin{equation}
c_{k-1} = u_{k}v_{k-1}-{(v_{k-1}\psi_{k})~'\over v_{k-1}}
\overline \psi_{k-1}+\psi_{k}\overline \psi_{k-1}~'
\label{4}
\end{equation}
takes place at each value of $k$,
where $c_k$ is an arbitrary constant. However, there is a simpler way
to check this statement by rewriting (\ref{4}) in the superfield form
\begin{equation}
c_{k-1} = \frac{D \overline D (f_k \overline f_{k-1} D \overline f_{k-1})}
{D \overline f_{k-1}}=
-\frac{\overline D D(f_k \overline f_{k-1}
\overline D f_{k})}{\overline D f_{k}}
\label{supcon}
\end{equation}
using eq. (\ref{2}) and definitions (\ref{com1}).
The last equality of formula (\ref{supcon}) demonstrates that
$c_k$ is a constant (i.e., it is the integral of motion), because it is
both the chiral and antichiral superfield, i.e. $Dc_k=\overline Dc_k=0$.
The action of the transformation $\sigma_m$ (\ref{auto1}) interchanges
different integrals in accordance with the following law:
\begin{equation}
\sigma_m c_k {\sigma}^{-1}_m= c_{m-k-1}.
\label{inter}
\end{equation}

Using the conservation law (\ref{4}), eq. (\ref{3}) for the fermionic
function $\psi_{k}$ can identically be rewritten as
\begin{equation}
({ (v_{k-1}\psi_{k})~'\over v_{k-1}})~'-
v_{k-1}[u_{k-1}-({\psi_{k-1}\overline \psi_{k-1}\over v_{k-1}})~'~]\psi_k=-
c_{k-1}\psi_{k-1}.
\label{5}
\end{equation}
 From eq. (\ref{5}), we observe that if we want to express
$\psi_k$ as a functional of the chain functions
$u_i$, $v_i$, $\psi_i$ and $\overline \psi_i$ defined at the previous
(from the left) points of the chain (i.e., at $i \leq k-1$), it is
necessary to solve a linear equation of the second order with
the coefficient functions that also depend on the functions of the
previous points of the chain. If we could find such an expression
for $\psi_k$, all other independent functions $\overline \psi_k$,
$u_k$ and $v_k$ of the f-Toda chain could also be obtained in a recurrent
way as functionals of the previous points of the chain. Indeed, using
eqs. (\ref{33}), (\ref{b2}) and conservation law (\ref{4}), one can
easily obtain the corresponding formulae for the functions $\overline
\psi_k$, $v_k$ and $u_k$:
\begin{eqnarray}
&& \overline \psi_k=
(-v_{k-1}({\overline \psi_{k-1}~'\over v_{k-1}})~'+
u_{k-1}\overline \psi_{k-1})/
(c_{k-1} + {(v_{k-1}\psi_{k})~'\over v_{k-1}}\overline \psi_{k-1}-
\psi_{k} {\overline \psi_{k-1}}'), \nonumber\\
&& v_{k}=v_{k-1}(-(\ln v_{k-1})''+u_{k-1}v_{k-1}+
{\psi_{k}}'\overline \psi_{k}-{\psi_{k-1}}'\overline \psi_{k-1})/
(c_{k-1}+{(\psi_{k} v_{k-1})~'\over v_{k-1}} \overline \psi_{k-1}-
\psi_{k} {\overline \psi_{k-1}}'), \nonumber\\
&& u_{k}=(c_{k-1}+{(\psi_{k} v_{k-1})~'\over v_{k-1}}\overline \psi_{k-1}-
\psi_{k} \overline \psi_{k-1}~')/v_{k-1},
\label{u}
\end{eqnarray}
respectively.

Thus, the crucial problem is the solution of eq. (\ref{5}) for
$\psi_k$ by some recurrent procedure. This task is solved in the next
subsection.

\subsection{The integrable factors}
The aim of this subsection is to construct the recurrent procedure for
solving eq. (\ref{5}) for an arbitrary chain point.

Let us assume that the homogeneous part of the second order equation
(\ref{5}) possesses a bosonic integrable factor $\mu_k$. This means
that after multiplication by $\mu_k$, the corresponding equation may be
represented as a whole derivative. Following this line, let us
rewrite eqs. (\ref{5}) in the following equivalent form:
\begin{equation}
(\frac{{\mu_{k}}^2}{v_{k-1}}({v_{k-1}\psi_{k}\over \mu_{k}})~'~)~'-
v_{k-1}\{-(\frac{\mu_k~'}{v_{k-1}})~'+ \mu_k [u_{k-1}-
({\psi_{k-1} \overline \psi_{k-1}\over v_{k-1}})~'~]\}\psi_k=-
c_{k-1}\mu_k \psi_{k-1}.
\label{m1}
\end{equation}
Equating the coefficient of the function $\psi_k$ in the l.h.s. of
eq.(\ref{m1}) to zero, we obtain the following equation:
\begin{equation}
({\mu_k~'\over  v_{k-1}})~'=[u_{k-1}-({\psi_{k-1}\overline \psi_{k-1}\over
v_{k-1}})~'~] \mu_k
\label{7}
\end{equation}
for the integrable factor $\mu_k$. At $k=1$, it has an obvious solution
$\mu_1=1$ because, in this case, its r.h.s. becomes equal to zero due to
the boundary conditions (\ref{bc1}).

If the integrable factors are restricted by eqs.
(\ref{7}), eq. (\ref{m1}) for the function $\psi_k$ becomes
\begin{equation}
(\frac{{\mu_{k}}^2}{v_{k-1}}({v_{k-1}\psi_{k}\over \mu_{k}})~'~)~'=-
c_{k-1}\mu_k \psi_{k-1}
\label{m2}
\end{equation}
and one can easily integrate it,
\begin{equation}
\psi_k={\mu_k\over v_{k-1}}[-c_{k-1}\int dz {v_{k-1}\over {\mu_k}^2}
\int dz \mu_k \psi_{k-1}+\beta_k\int dz {v_{k-1}\over {\mu_k}^2}+\alpha_k],
\label{10}
\end{equation}
where $\alpha_k$ and $\beta_k$ are fermionic constants of integration,
and one can find the explicit expression for $\psi_k$ if all other
parts of formula (\ref{10}) are known. The appearance of
these additional constants is a qualitative difference of the f-Toda
chain in comparison with the integrable chains investigated before.

Concerning eqs. (\ref{7}) for the factors $\mu_k$, they
can be resolved in recurrent form with the following answer:
\begin{equation}
\mu_{k+1}=v_{k-1}(\frac{\mu_{k}}{v_{k-1}})~'-
\frac{{\mu_{k}}^2}{c_{k-1}}({v_{k-1}\psi_{k}\over \mu_{k}})~'~
(\frac{\overline \psi_{k-1}}{v_{k-1}})~', \quad
\mu_{1}=1.
\label{rec}
\end{equation}
Indeed, this can be checked by direct substitution, taking into account
the f-Toda chain equations (\ref{b1})-(\ref{b2}). If the integrable
factor $\mu_k$ satisfies eq.(\ref{7}), the factor $\mu_{k+1}$ (\ref{rec})
satisfies the equation
\begin{equation}
({\mu_{k+1}~'\over  v_{k}})~'=[u_{k}-
({\psi_{k}\overline \psi_{k}\over v_{k}})~'~] \mu_{k+1}.
\label{mu}
\end{equation}
This verification can be simplified if one uses the following
convenient identity:
\begin{eqnarray}
(\frac{1}{v_{k}}(v_{k-1}(\frac{\overline
\psi_{k-1}}{v_{k-1}})~'~)~'~)~'=
[u_{k}-({\psi_{k}\overline \psi_{k}\over v_{k}})~'~]~
v_{k-1} (\frac{\overline \psi_{k-1}}{v_{k-1}})~'
-c_{k-1}({\overline \psi_{k}\over v_{k}})~',
\label{ident}
\end{eqnarray}
which can easily be checked.

Let us stress that the integrable factor $\mu_{k+1}$ (\ref{rec}) is the
functional of the chain functions defined at the previous (from the left)
points of the chain, thus, the same important property is also satisfied
for the function $\psi_k$ (\ref{10}).

To consider the case of the f-Toda chain interrupted from the right with
the boundary conditions (\ref{bc2}), one can simply apply the
transformation $\sigma_M$ (\ref{auto1})
to formulae (\ref{5})--(\ref{ident}). Without additional
comments, let us present expressions for $\overline \psi_k$ and their
integrable factors ${\overline \mu_k}$
\begin{equation}
\overline \psi_k={\overline \mu_k\over u_{k+1}}[-c_{k}\int dz {u_{k+1}\over
{\overline \mu_k}^2} \int dz \overline \mu_k \overline
\psi_{k+1}+\overline \beta_k\int dz {u_{k+1}\over {{\overline
\mu}_k}^2}+\overline \alpha_k],
\label{10bar}
\end{equation}
\begin{equation}
\overline \mu_{k-1}=u_{k+1}(\frac{\overline \mu_{k}}{u_{k+1}})~'+
\frac{{\overline \mu_{k}}^2}{c_{k}}(\frac{\psi_{k+1}}{u_{k+1}})~'~
({u_{k+1}\overline \psi_{k}\over \overline \mu_{k}})~', \quad
\overline \mu_{M-1}=1,
\label{recbar}
\end{equation}
where $\overline \alpha_k$ and $\overline \beta_k$ are fermionic constants,
which will be useful in what follows.

Thus, the following proposition summarizes this section.

{\it The solution of the f-Toda chain interrupted from the left
is given by recurrent relations (\ref{10}), (\ref{rec}) and (\ref{u}).}

\section{The integrability of the f-Toda chain with fixed ends}
In this section, we construct the recurrent procedure for solving
the f-Toda chain with the boundary conditions (\ref{bc1}),
(\ref{bc2}).

\subsection{New form of the f-Toda chain}
Let us introduce the new basis $q_k$, $r_k$, $\xi_k$ and $\overline \xi_k$
in the space of the f-Toda chain functions $u_k$, $v_k$, $\psi_k$ and
$\overline \psi_k$, defined by the following invertible transformation:
\begin{eqnarray}
&& q_k = \prod_{s=1}^k (u_{s}v_{s-1}),\quad
r_0=v_0, \quad r_k=v_k \prod_{s=1}^k (u_{s}v_{s-1}), \nonumber\\
&& \overline \xi_0= \overline \psi_0, \quad
\overline \xi_k= \overline \psi_k \prod_{s=1}^k (u_{s}v_{s-1}), \quad
\xi_k= {\psi_k}/{\prod_{s=1}^k (u_{s}v_{s-1})},
\label{trans1}
\end{eqnarray}
\begin{eqnarray}
u_k = \frac{q_k}{r_{k-1}}, \quad v_0=r_0, \quad v_k=\frac{r_k}{q_k}, \quad
\overline \psi_0 = \overline \xi_0, \quad
\overline \psi_k = \frac{\overline \xi_k}{q_k}, \quad
\psi_k=\xi_k q_k,
\label{trans2}
\end{eqnarray}
where $k=1,2,...,M$. In this case, the boundary conditions (\ref{bc1}) and
(\ref{bc2}) are transformed into the following conditions:
\begin{equation}
q_0=\xi_0=0,
\label{bcn1}
\end{equation}
\begin{equation}
r_M=\overline \xi_M=0,
\label{bcn2}
\end{equation}
respectively. In the new basis, the f-Toda chain equations
(\ref{b1})-(\ref{b2}) have the following form:
\begin{equation}
(\ln q_{j})~'=\xi_{j}\overline \xi_{j}, \quad
(\ln q_{M})~'=0,
\label{eqb1}
\end{equation}
\begin{eqnarray}
-(r_{M-1} \xi_{M}~')~'=r_{M-1}\xi_{M-1}, \quad
-({\overline \xi_0~'\over r_0})~'={\overline \xi_{1}\over r_0},
\label{eqf1}
\end{eqnarray}
\begin{eqnarray}
-(r_{j-1} \xi_{j}~')~'=-r_{j}\xi_{j}+r_{j-1}\xi_{j-1},
\label{eq1}
\end{eqnarray}
\begin{eqnarray}
-({\overline \xi_j~'\over r_j})~'={\overline \xi_{j+1}\over r_j}-
{\overline \xi_j\over r_{j-1}},
\label{eq}
\end{eqnarray}
\begin{eqnarray}
-(\ln r_0)~''={r_{1}\over r_0}-\xi_{1}~'~\overline \xi_{1}, \quad
-(\ln r_j)~''={r_{j+1}\over r_j}-{r_j\over r_{j-1}}-\xi_{j+1}~'~
\overline \xi_{j+1} -\xi_j \overline \xi_j~'.
\label{eqb2}
\end{eqnarray}
Hereafter, the index $j$ lies in the following range $1 \leq j \leq M-1$.

The substitution of the transformations (\ref{trans1}), (\ref{trans2}) into
(\ref{auto1}) at $m=M$ gives the inner automorphism $\sigma_M$,
\begin{eqnarray}
\sigma_M q_k {\sigma}^{-1}_M= \frac{q_M}{q_{M-k}} , \quad
\sigma_M r_k {\sigma}^{-1}_M= \frac{q_M}{r_{M-k-1}}, \quad
\sigma_M \xi_k {\sigma}^{-1}_M=\frac{\overline \xi_{M-k}}{q_M}, \quad
\sigma_M {\overline \xi_k} {\sigma}^{-1}_M=q_M \xi_{M-k}, ~~~~~
\label{auto2}
\end{eqnarray}
for eqs. (\ref{eqb1})--(\ref{eqb2}).

It is interesting to note that the functions $q_k$, $\xi_{M}$ and
$\overline \xi_0$ are completely decoupled from eqs. (\ref{eq1}),
(\ref{eq}) and (\ref{eqb2}), which form a closed set of equations for
the functions $r_0$, $r_j$ $\xi_j$ and $\overline \xi_j$. Moreover, taking
into account the corollary from eq. (\ref{eqb1}) that $q_M$ is an
arbitrary constant, one can conclude that the transformations
(\ref{auto2}) are also closed for them. However, this is not the case for
the $N=2$ supersymmetry transformations (\ref{suptrans}), which, for
eqs. (\ref{eqb1})--(\ref{eqb2}), have the following infinitesimal
form:
\begin{eqnarray}
&& \delta q_k  =- \prod_{s=1}^k ( \overline \varepsilon \xi_s~'r_{s-1}+
\varepsilon \frac{\overline \xi_s~'}{r_{s-1}})\frac{q_s}{q_{s-1}},
\nonumber\\
&&\delta r_k  =- \varepsilon \overline \xi_k~'+
\frac{q_k \delta q_k}{r_{k-1}}, \quad
\delta \overline \xi_k  =\overline \varepsilon r_k+
\frac{\overline \xi_k \delta q_k}{q_{k}}, \quad
\delta \xi_k  = \frac{\varepsilon}{r_{k-1}}-
\frac{\xi_k \delta q_k}{q_{k}},
\label{suptrans1}
\end{eqnarray}
where eqs. (\ref{eqb1}) have been used.

We call eqs. (\ref{eq1})--(\ref{eqb2}) the restricted f-Toda chain.
If some of their explicit solutions are known, one can also easily obtain
the solutions of eqs. (\ref{eqb1})--(\ref{eqf1}) for $q_j$, $\xi_{M}$ and
$\overline \xi_0$,
\begin{eqnarray}
q_{j+1} = c_{j}q_{j}/
(1-{(r_{j}\xi_{j+1})~'\over r_{j}} \overline \xi_{j}+
\xi_{j+1}\overline \xi_{j}~'), \quad
q_{1} = c_{0}/(1-{(r_{0}\xi_{1})~'\over r_{0}} \overline \xi_{0}+
\xi_{1}\overline \xi_{0}~'),
\label{q}
\end{eqnarray}
\begin{eqnarray}
&& \xi_M = -\int dz {1 \over r_{M-1}} \int dz r_{M-1} \xi_{M-1} +
\beta_M \int dz {1\over r_{M-1}} + \alpha_M, \nonumber\\
&& \overline \xi_0 = -\int dz r_{0} \int dz {\overline \xi_{1}\over r_{0}}
+ \overline \beta_0 \int dz r_{0} + \overline \alpha_0,
\label{cons}
\end{eqnarray}
where $\alpha_M,\overline \alpha_0$ and $\beta_M,\overline \beta_0$ are
the fermionic constants of integration and we have used the conservation
laws (\ref{4}) expressed in terms of the new functions $q_k$, $r_k$,
$\xi_k$ and $\overline \xi_k$ (\ref{trans1}), (\ref{trans2}).

Let us introduce the new bosonic function $x_j$
\begin{eqnarray}
r_0=\exp (x_0), \quad r_j=\exp (x_j), \quad
\exp(-x_{-1})=\exp(x_M)=0,
\label{r}
\end{eqnarray}
which is usually used in the case of
the bosonic Toda chain and is more suitable for the system under
consideration. In this case, the restricted f-Toda chain
equations (\ref{eq1})--(\ref{eqb2}) and their inner automorphism $\sigma_M$
become
\begin{eqnarray}
\xi_{j}~''+x_{j-1}~'~ \xi_{j}~'-\xi_{j}\exp(x_{j}-x_{j-1})=-\xi_{j-1},
\label{eqn1}
\end{eqnarray}
\begin{eqnarray}
\overline \xi_j~''- x_j~'~\overline \xi_j~'-
\overline \xi_j \exp(x_{j}-x_{j-1})=-\overline \xi_{j+1} ,
\label{eqn}
\end{eqnarray}
\begin{eqnarray}
&& x_0~''+\exp(x_1-x_0)-\xi_{1}~'~\overline \xi_{1}=0, \nonumber\\
&& x_j~''+\exp(x_{j+1}- x_j)-\exp(x_j- x_{j-1})-\xi_{j+1}~'~
\overline \xi_{j+1} -\xi_j \overline \xi_j~'=0,
\label{eqbn2}
\end{eqnarray}
\begin{eqnarray}
\sigma_M x_0 {\sigma}^{-1}_M= -x_{M-1}, \quad
\sigma_M x_j {\sigma}^{-1}_M= -x_{M-j-1}, \quad
\sigma_M \xi_j {\sigma}^{-1}_M=\overline \xi_{M-j}, \quad
\sigma_M {\overline \xi_j} {\sigma}^{-1}_M= \xi_{M-j}, ~ ~ ~ ~
\label{n0}
\end{eqnarray}
respectively. In what follows, we concentrate on the analysis of these
equations and prove their integrability.

\subsection{The general assertion}
In this subsection, we briefly discuss the general problem of
integrating a system of ordinary differential equations containing both
unknown fermionic and bosonic functions.

Althought this is not crucial for the general discussion, for
definiteness, we assume that the system consists of the second order
differential equations for $M$ bosonic and $2(M-1)$ fermionic independent
functions, as takes place for eqs. (\ref{eqn1})--(\ref{eqbn2}). In
this case, its general exact solution must include $2M$ bosonic and
$4(M-1)$ fermionic independent constants, of course, if the system is
integrable.

Let us analyze the qualitative structure of this exact solution in more
detail.

Evidently, the exact solution is a polynomial with respect to
fermionic constants for any function involved in the system. It is well
known that the bosonic (fermionic) functions are even (odd) polynomials
and any fermionic constant enters a monomial representing the
product of fermionic components only once\footnote{Let us
recall that each polynomial is the sum of monomials composed of the
products of different fermionic components and, by definition,
its degree is the number of components of its maximum monomial.}.

It is instructive to treat this exact solution as the result of some
iteration procedure or as a result of calculations in the framework of
some perturbation theory, where the role of perturbation parameters is
played by the fermionic constants, and different orders of perturbation
calculations are in one-to-one correspondence with different degrees
of the monomials, i.e., with the different numbers of fermionic components
composing them.

Now we will analyze the qualitative structure of different orders of
perturbation calculations in the framework of such a perturbation theory.

The zero-order approximation to the exact solution corresponds to the
general solution of a pure bosonic system, which can be derived by taking
the bosonic limit of the initial system. All of the bosonic integration
constants parametrising the exact solution of the initial system must
arise in this order of perturbative theory.

The first order approximation corresponds to the general solution of a
linear fermionic system, which can be derived by substituting the
zero-order approximation for bosonic functions into the initial system
linearized with respect to fermionic functions. Similarly to the case
of zero-order approximation, all fermionic integration constants
parametrising the exact solution of the initial system must arise in this
order of perturbative theory.

The second-order correction corresponds to the solution of a
linear bosonic system with the following structure: its homogeneous part
is the symmetry equation$^2$ of a zero-order bosonic system and
its inhomogeneous part is what remains after the substitution of the zero
and first orders of perturbative calculations into the initial system
and rejection of the monomials of all degrees but monomials of the
second degree.

The homogeneous part of a fermionic system, corresponding to the third
order-correction to the exact solution, coincides with the fermionic
system of the first order-approximation and its inhomogeneous part is
given by perturbative decomposition of the initial system using the
functions of the first and second orders of the perturbative calculations,
and so on.

To close this general discussion, let us stress that the homogeneous part
of a system corresponding to any even (odd) order-correction to the exact
solution coincides with the system of the second (first) order-approximation
and neither new bosonic nor fermionic independent parameters appear
in any order starting with the second order of the
perturbation calculations\footnote{Clearly, having arisen, they can always
be eliminated by redefining the parameters of the zero and first orders of
the perturbation calculations.}. We would also like to emphasize
that by construction, such a perturbation theory is convergent and it
gives the exact solution: the perturbation series is interrupted due to
the fermionic nature of the perturbative parameters because a limited
number of monomials can be composed using a finite set of fermionic
constants (i.e., the solutions of the first-order approximation).

Now let us formulate the assertion with respect to the initial system:

{\it if the equations of the zero and first orders of the perturbation
theory are exactly integrable, the initial system
is also integrable, at least in the quadratures}.

The proof is given in a few words.

If the equations of the zero (first) order of the perturbative theory
are exactly integrable, the corresponding symmetry equation is
also exactly integrable. Indeed, one can derive its $2M$ ($4(M-1)$)
linear-independent solutions by taking the derivatives of the general
solution of the zero (first) order equations with respect to its $2M$
($4(M-1)$) bosonic (fermionic) independent constants.
As follows from the previous discussion, the problem of
resolving equations corresponding to any other order of the
perturbative calculations reduces to the problem of resolving the
inhomogeneous symmetry equation, but the last problem is an exactly
solvable one. Indeed, if the general solution of some linear homogeneous
system is known, like in the case under consideration, then, by applying
the well-known method of varying the arbitrary constants, one can
algorithmically construct the solution of the corresponding inhomogeneous
equation, at least, in the quadratures. Thus, all the orders of the
above-discussed perturbation calculations can be resolved in explicit form.

\subsection{Solution of the restricted f-Toda chain}
The purpose of this subsection is to apply the above-developed regular
algorithm of integrating a system of ordinary differential equations
containing both unknown fermionic and bosonic functions to the restricted
f-Toda chain equations (\ref{eqn1})--(\ref{eqbn2}).

Following the line of section 5.2, let us represent the functions
$x_0$, $x_j$, $\xi_j$ and $\overline \xi_j$ as a perturbation series,
\begin{eqnarray}
&& x_0=\sum_{l=0}^{2(M-1)} x^{(2l)}_0, \quad
x_j=\sum_{l=0}^{2(M-1)} x^{(2l)}_j, \quad
\xi_j=\sum_{l=1}^{2(M-1)} \xi^{(2l-1)}_j, \quad
\overline \xi_j=\sum_{l=1}^{2(M-1)} \overline \xi^{(2l-1)}_j,
\label{pert}
\end{eqnarray}
where the perturbation parameters coincide with the $4(M-1)$ fermionic
constants of the first-order approximation to the general solution. Here,
the functions $x^{(2l)}_0$ and $x^{(2l)}_j$ ($\xi^{(2l+1)}_j$ and
$\overline \xi^{(2l+1)}_j$) are the $2l$ ($2l+1$) order
corrections to the zero (first) order approximation. The fermionic
character of the decomposition parameters guarantees that this series is
interrupted, starting with the $4(M-1)+1$ order. As result, such a
perturbation theory is convergent and gives an exact result.

Substituting decompositions (\ref{pert}) into
eqs. (\ref{eqn1})--(\ref{eqbn2}), extracting terms of the same
order, and equating their algebraic sum to zero independently for
different orders, one can obtain the complete set of perturbative
equations corresponding to the presented perturbation theory.

Here, we demonstrate that the conditions of the general
assertion in the previous subsection is satisfied for the restricted
f-Toda chain, i.e., it is exactly integrable.

The zero order bosonic system coincides with the usual one-dimensional
Toda chain \cite{t}
\begin{eqnarray}
&& x^{(0)}_0~''+\exp(x^{(0)}_1-x^{(0)}_0)=0, \nonumber\\
&& x^{(0)}_j~''+\exp(x^{(0)}_{j+1}- x^{(0)}_j)-\exp(x^{(0)}_j-
x^{(0)}_{j-1})=0,
\label{toda}
\end{eqnarray}
which is exactly integrable. Its general solution is well
known and can be represented in the following form \cite{l}:
\begin{eqnarray}
r^{(0)}_0 = \exp (x^{(0)}_0)=\sum_{i=1}^{M} a_i \exp (b_i x), \quad
\exp (x^{(0)}_j)=(-1)^j {Det_{j+1}\over Det_j},
\label{sol}
\end{eqnarray}
where $a_i$ and $b_i$ are arbitrary constants, and $Det_j$ is the $j$th
principal minor of the matrix
\begin{eqnarray}
\pmatrix{r^{(0)}_0    & r^{(0)}_0~'     & r^{(0)}_0~''   & \ldots \cr
         r^{(0)}_0~'  & r^{(0)}_0~''    & r^{(0)}_0~'''  & \ldots \cr
	 r^{(0)}_0~'' & r^{(0)}_0~'''   & r^{(0)}_0~'''' & \ldots \cr
	  \vdots      & \vdots          & \vdots         &         \cr }.
\end{eqnarray}

The first order fermionic systems have the following form:
\begin{eqnarray}
\xi^{(1)}_{j}~''+x^{(0)}_{j-1}~'~ \xi^{(1)}_{j}~'-
\xi^{(1)}_{j}\exp(x^{(0)}_{j}-x^{(0)}_{j-1})+\xi^{(1)}_{j-1}=0,
\label{eqn1z}
\end{eqnarray}
\begin{eqnarray}
\overline \xi^{(1)}_j~''- x^{(0)}_j~'~\overline \xi^{(1)}_j~'-
\overline \xi^{(1)}_j \exp(x^{(0)}_{j}-x^{(0)}_{j-1})+
\overline \xi^{(1)}_{j+1}=0.
\label{eqnz}
\end{eqnarray}
These equations are also exactly integrable and
using the results of section 4.2, one can easily generate their general
solutions. Thus, substituting transformations (\ref{trans2}), (\ref{r}),
as well as solutions (\ref{q}), into relations (\ref{10}),
(\ref{rec}), (\ref{10bar}), and (\ref{recbar}), and rejecting terms
nonlinear with respect to the fermionic fields, we get the following
recurrent formulae\footnote{Here, we have rescaled
the fermionic constants of expressions (\ref{10}) and (\ref{10bar}).}:
\begin{eqnarray}
&& \xi^{(1)}_j=\exp(-x^{(0)}_{j-1})\mu^{(0)}_j
[-\int dz {\exp(x^{(0)}_{j-1})\over {\mu^{(0)}_j}^2}
\int dz \mu^{(0)}_j \xi^{(1)}_{j-1}+\beta_j\int dz {\exp(x^{(0)}_{j-1})\over
{\mu^{(0)}_j}^2}+\alpha_j], \nonumber\\
&& \mu^{(0)}_{j+1}=\mu^{(0)}_{j}~'-x^{(0)}_{j-1}~'~\mu^{(0)}_{j},
\quad \mu^{(0)}_{1}=1,
\label{mm1}
\end{eqnarray}
\begin{eqnarray}
&& \overline \xi^{(1)}_j=\exp(x^{(0)}_{j})\overline \mu^{(0)}_j
[-\int dz {\exp(-x^{(0)}_{j})\over{\overline \mu^{(0)}_j}^2}
\int dz \overline \mu^{(0)}_j~\overline \xi^{(1)}_{j+1}+
\overline \beta_j\int dz {\exp(-x^{(0)}_{j})\over
{{\overline\mu}^{(0)}_j}^2}+\overline \alpha_j], \nonumber\\
&& \overline \mu^{(0)}_{j-1}=\mu^{(0)}_{j}~'+
x^{(0)}_{j}~'~\overline \mu^{(0)}_{j}, \quad
\overline \mu^{(0)}_{M-1}=1
\label{mm3}
\end{eqnarray}
for the exact solutions of eqs. (\ref{eqn1z})--(\ref{eqnz}).
Thus, following the general assertion of section 5.2, we conclude
that the restricted f-Toda chain is also integrable.

Now we show how to construct the solutions of the perturbation equations
for all other orders of the perturbation theory with respect to the
$4(M-1)$ fermionic constants $\alpha_j$, $\beta_j$, $\overline \alpha_j$,
and $\overline \beta_j$ of the first order solutions (\ref{mm1}),
(\ref{mm3}).

For the perturbation equations of the $2l$ $(l \geq 1)$ and $2l-1$
$(l \geq 2)$ orders, we have
\begin{eqnarray}
&& x^{(2l)}_0~''+(x^{(2l)}_{1}-x^{(2l)}_0)\exp (x^{(0)}_{1}-x^{(0)}_0)=
X^{(2l)}_0, \nonumber\\
&& x^{(2l)}_j~''+(x^{(2l)}_{j+1}-x^{(2l)}_j)\exp (x^{(0)}_{j+1}-x^{(0)}_j)-
(x^{(2l)}_j-x^{(2l)}_{j-1})\exp (x^{(0)}_j-x^{(0)}_{j-1})=X^{(2l)}_j,
\label{2lord}
\end{eqnarray}
\begin{eqnarray}
\xi^{(2l-1)}_{j}~''+x^{(0)}_{j-1}~'~ \xi^{(2l-1)}_{j}~'-
\xi^{(2l-1)}_{j}\exp(x^{(0)}_{j}-x^{(0)}_{j-1})+\xi^{(2l-1)}_{j-1}=
\Xi^{(2l-1)}_j,
\label{2l1ord1}
\end{eqnarray}
\begin{eqnarray}
\overline \xi^{(2l-1)}_j~''- x^{(0)}_j~'~\overline \xi^{(2l-1)}_j~'-
\overline \xi^{(2l-1)}_j \exp(x^{(0)}_{j}-x^{(0)}_{j-1})+
\overline \xi^{(2l-1)}_{j+1}=\overline \Xi^{(2l-1)}_j,
\label{2l1ord2}
\end{eqnarray}
respectively. Here, the functions $X^{(2l)}_0$ and $X^{(2l)}_j$
($\Xi^{(2l-1)}_j$ and $\overline \Xi^{(2l-1)}_j$)
are defined by the following relations:
\begin{eqnarray}
X_0 \equiv &&
\sum_{l=1}^{2(M-1)}X^{(2l)}_0=
-\exp(x_1-x_0)+\xi_{1}~'~\overline \xi_{1}+
(1+x_1-x_0-x^{(0)}_{1}+x^{(0)}_0)\exp (x^{(0)}_{1}-x^{(0)}_0), \nonumber\\
X_j \equiv &&
\sum_{l=1}^{2(M-1)}X^{(2l)}_j=
-\exp(x_{j+1}- x_j)+\exp(x_j- x_{j-1})+
\xi_{j+1}~'~ \overline \xi_{j+1}+ \xi_j~ \overline \xi_j~' \nonumber\\
&& +(1+x_{j+1}-x_{j}-x^{(0)}_{j+1}+x^{(0)}_j)\exp (x^{(0)}_{j+1}-x^{(0)}_j)
\nonumber\\
&&-(1+x_{j}-x_{j-1}-x^{(0)}_{j}+x^{(0)}_{j-1})
\exp (x^{(0)}_{j}-x^{(0)}_{j-1}), \nonumber\\
\Xi_j \equiv &&
\sum_{l=2}^{2(M-1)}\Xi^{(2l-1)}_{j}=
(x^{(0)}_{j-1}-x_{j-1})~'~ \xi_{j}~'+
\xi_{j}(\exp(x_{j}-x_{j-1})-\exp(x^{(0)}_{j}-x^{(0)}_{j-1})), \nonumber\\
\overline \Xi_j \equiv &&
\sum_{l=2}^{2(M-1)}\overline \Xi^{(2l-1)}_j=
-(x^{(0)}_{j}- x_j)~'~\overline \xi_j~'+
\overline \xi_j (\exp(x_{j}-x_{j-1})-\exp(x^{(0)}_{j}-x^{(0)}_{j-1})),
\label{monoms}
\end{eqnarray}
respectively, and they are the sum of the monomials of $2l$ ($2l-1$)
degree\footnote{It is implied that the perturbative decompositions
(\ref{pert}) of the functions $x_0$, $x_j$, $\xi_j$ and $\overline \xi_j$
must be substituted into the r.h.s. of relations (\ref{monoms}).}.
Let us stress that the inhomogeneous parts of perturbative equations
(\ref{2lord})-(\ref{2l1ord2}), corresponding to a given order $n$,
depend only on the functions $x^{(2m)}_0$,
$x^{(2m)}_j$, $\xi^{(2l+1)}_j$ and $\overline \xi^{(2l+1)}_j$ of the
previous orders of perturbative calculations (i.e., at
$m \leq (n-1)/2$ and $l \leq (n-2)/2$). Thus, we have a consistent
perturbative theory.

If we set the inhomogeneous parts of perturbative equations (\ref{2lord})
equal to zero for any even order of the perturbative theory, they coincide
with the symmetry equations$^2$ for the Toda chain (\ref{toda}).
Thus, in this case, their $2M$ linear--independent solutions
$y^{\Lambda}_k$ at each point $k$ ($k=0,1, ... ,M-1$) of the chain are
given by
\begin{eqnarray}
y^{\Lambda}_k=\frac{\partial x^{(0)}_k}{\partial
A_{\Lambda}},
\end{eqnarray}
where the capital Greek letter-indices run
over the range $\Lambda, \Phi = 0,1,...,2M-1$, and $A_{\Lambda} \equiv
\{a_1,...,a_M, b_1,...,b_M\}$ (see eq. (\ref{sol})). To use the
formalism of varied constants for the solution of inhomogeneous
equations (\ref{2lord}), it is necessary to solve the linear system of the
first order differential equations
\begin{eqnarray}
\sum_{{\Lambda}=0}^{2M-1} y^{\Lambda}_k c^{(2l)}_{\Lambda}(z)~'=0 , \quad
\sum_{{\Lambda}=0}^{2M-1} y^{\Lambda}_k~'~ c^{(2l)}_{\Lambda}(z)~'=X^{(2l)}_k
\label{bbb}
\end{eqnarray}
with respect to $c^{(2l)}_{\Lambda}(z)$, where
$c^{(2l)}_{\Lambda}(z)$ are $2M$ bosonic parameter-functions. Then, the
solution of eqs. (\ref{2lord}) for the function $x^{(2l)}_k$ has the
following form:
\begin{eqnarray}
x^{(2l)}_k=\sum_{{\Lambda}=0}^{2M-1} y^{\Lambda}_k c^{(2l)}_{\Lambda}(z).
\label{intb}
\end{eqnarray}

Let us introduce the $2M \times 2M$ matrix ${\cal P}_{\Phi}^{\Lambda}$
defined as
\begin{eqnarray}
{\cal P}_{k}^{\Lambda}=y^{\Lambda}_k~', \quad
{\cal P}_{M+k}^{\Lambda}=y^{\Lambda}_k.
\label{matrp}
\end{eqnarray}
Then the solution of the system (\ref{bbb}) can be represented in the form
\begin{eqnarray}
c^{(2l)}_{\Lambda}(z)=
\int dz \sum_{m=0}^{M-1} ({\cal P}^{-1})_{\Lambda}^m X^{(2l)}_m,
\label{bbbb}
\end{eqnarray}
where ${\cal P}^{-1}$ is the inverse matrix for the matrix
${\cal P}$ (${\cal P}^{-1}{\cal P}={\cal P}{\cal P}^{-1}=I$)
and expression (\ref{intb}) for $x^{(2l)}_k$ becomes
\begin{eqnarray}
x^{(2l)}_k=\sum_{{\Lambda}=0}^{2M-1} {\cal P}_{M+k}^{\Lambda}
\int dz \sum_{m=0}^{M-1} ({\cal P}^{-1})_{\Lambda}^m
X^{(2l)}_m.
\label{intbb}
\end{eqnarray}
Taking the sum over all orders of the perturbative theory and using
relations (\ref{pert}) and (\ref{monoms}), we have the following exact
expression for $x_k$:
\begin{eqnarray}
x_k= x_k^{(0)}+\sum_{{\Lambda}=0}^{2M-1} {\cal P}_{M+k}^{\Lambda}
\int dz \sum_{m=0}^{M-1}
({\cal P}^{-1})_{\Lambda}^m X_m.
\label{intbbb}
\end{eqnarray}
One can invert the above-developed perturbation scheme and consider the
relation (\ref{intbbb}) as the equation for $x_k$, and take it
as a starting point. Then its iteration (\ref{pert}), together with
the iteration of the corresponding equations for the fermionic functions
$\xi_j$ and $\overline \xi_j$ (see below), is interrupted starting with
the $4(M-1)+1$ order and gives the exact solution for $x_k$.

The odd-order perturbative equations (\ref{2l1ord1}), (\ref{2l1ord2}) can
be integrated in the same way as described for the case of the
even-order equations (\ref{2lord}). Without going into detail, let us
present only the expressions for the functions $\xi^{(2l-1)}_j$ and
$\overline \xi^{(2l-1)}_j$,
\begin{eqnarray}
\xi^{(2l-1)}_j=\sum_{{\Omega}=1}^{2(M-1)}
\frac{\partial \xi^{(1)}_j}{\partial \Gamma_{\Omega}}c^{(2l-1)}_{\Omega}(z),
\quad \overline \xi^{(2l-1)}_j=\sum_{{\Omega}=1}^{2(M-1)}
\frac{\partial \overline \xi^{(1)}_j}{\partial
\overline \Gamma_{\Omega}}\overline c^{(2l-1)}_{\Omega}(z),
\label{intf}
\end{eqnarray}
where $\Gamma_{\Omega}=(\alpha_1,...,\alpha_{M-1},\beta_1,...,\beta_{M-1})$,
${\overline \Gamma}_{\Omega}=(\overline \alpha_1,...,\overline \alpha_{M-1},
\overline \beta_1,...,\overline \beta_{M-1})$ ($\Omega=1,2,...,2(M-1)$);
$c^{(2l-1)}_{\Omega}(z)$ and $\overline c^{(2l-1)}_{\Omega}(z)$
are $4(M-1)$ fermionic parameter-functions that are solutions of the
following linear system:
\begin{eqnarray}
&& \sum_{{\Omega}=1}^{2(M-1)}
\frac{\partial \xi^{(1)}_j}{\partial \Gamma_{\Omega}}
c^{(2l-1)}_{\Omega}(z)~'=0 , \quad
\sum_{{\Omega}=1}^{2(M-1)}
(\frac{\partial \xi^{(1)}_j}{\partial \Gamma_{\Omega}})~'~
c^{(2l-1)}_{\Omega}(z)~'=\Xi^{(2l-1)}_j , \nonumber\\
&& \sum_{{\Omega}=1}^{2(M-1)}
\frac{\partial \overline \xi^{(1)}_j}{\partial \overline \Gamma_{\Omega}}
\overline c^{(2l-1)}_{\Omega}(z)~'=0 , \quad
\sum_{{\Omega}=1}^{2(M-1)}
(\frac{\partial \overline \xi^{(1)}_j}{\partial
\overline \Gamma_{\Omega}})~'~\overline c^{(2l-1)}_{\Omega}(z)~'=
\overline \Xi^{(2l-1)}_j.
\label{fff}
\end{eqnarray}

To close this subsection, we would like briefly discuss a slightly modified
version of the perturbative theory for the odd-order perturbative
equations (\ref{2l1ord1}) and (\ref{2l1ord2}).

To do this, let us represent the differential equations (\ref{2l1ord1})
and (\ref{2l1ord2}) in integral form. Thus, multiplying them by the
integrable factors $\mu^{(0)}_j$ and $\overline \mu^{(0)}_j$,
respectively, using the equations
\begin{eqnarray}
(\exp(-x^{(0)}_{j-1})~
\mu^{(0)}_j~'~)~'=\exp(-x^{(0)}_{j-2})~ \mu^{(0)}_j, \quad
(\exp(x^{(0)}_{j})~ \overline \mu^{(0)}_j~'~)~'=\exp(x^{(0)}_{j-1})~
\overline \mu^{(0)}_j
\label{n1}
\end{eqnarray}
for the factors, taking the sum of all perturbative orders, and using
relations (\ref{monoms}), eqs. (\ref{2l1ord1})--(\ref{2l1ord2}) can be
identically rewritten in the form
\begin{eqnarray}
&& (\exp(-x^{(0)}_{j-1})~ {\mu^{(0)}_{j}}^2~
({\exp(x^{(0)}_{j-1})\over \mu^{(0)}_{j}}\xi_{j})~'~)~'=
\mu^{(0)}_j (-\xi_{j-1} + \Xi_j),
\nonumber\\
&& (\exp(x^{(0)}_{j})~ {\overline \mu^{(0)}_{j}}^2~
({\exp(-x^{(0)}_{j})\over \overline \mu^{(0)}_{j}}\overline \xi_{j})~'~)~'=
\overline \mu^{(0)}_j (-\overline \xi_{j+1} + \overline \Xi_j)
\label{nn1}
\end{eqnarray}
which can be easily integrated. As a result, we have the desirable
integral form of eqs. (\ref{2l1ord1})--(\ref{2l1ord2}) given by
\begin{eqnarray}
&& \xi_j=\exp(-x^{(0)}_{j-1}) \mu^{(0)}_j
[\int dz {\exp(x^{(0)}_{j-1})\over {\mu^{(0)}_j}^2}\int dz \mu^{(0)}_j
(-\xi_{j-1}+ \Xi_j)+
\beta_j\int dz {\exp(x^{(0)}_{j-1})\over {\mu^{(0)}_j}^2}+\alpha_j], ~ ~ ~
\nonumber\\
&& \overline \xi_j=\exp(x^{(0)}_{j}) \overline \mu^{(0)}_j
[\int dz {\exp(-x^{(0)}_{j})\over {\overline \mu^{(0)}_j}^2}
\int dz \overline \mu^{(0)}_j
(-\overline \xi_{j+1}+ \overline \Xi_j)+
\overline \beta_j\int dz {\exp(-x^{(0)}_{j})\over
{\overline \mu^{(0)}_j}^2}+\overline \alpha_j]. ~ ~ ~
\label{n4}
\end{eqnarray}
Simple inspection of eqs. (\ref{n4}) shows that their solutions can be
consistently obtained by iterating with respect to the fermionic
constants of integration $\alpha_j$, $\beta_j$, $\overline \alpha_j$ and
$\overline \beta_j$ in the framework of the above-discussed
perturbative scheme.

Thus, all the orders of the perturbation calculations can be resolved in
explicit form.

\subsection{Example: the $M=2$ case}

To illustrate the general formulae of the previous subsection,
here we consider the simplest example of the restricted f-Toda chain
with fixed ends at $M=2$.

In this case, eqs. (\ref{eqn1})--(\ref{eqbn2}), boundary conditions
(\ref{bcn1}), (\ref{bcn2}), (\ref{r}), as well as perturbative
decompositions (\ref{pert}), have the following form:
\begin{eqnarray}
&& \xi_{1}~''+x_{0}~'~
\xi_{1}~'-\xi_{1}\exp(x_{1}-x_{0})=0, \quad \overline \xi_1~''-
x_1~'~\overline \xi_1~'- \overline \xi_1 \exp(x_{1}-x_{0})=0 , \nonumber\\
&& x_0~''+\exp(x_1-x_0)-\xi_{1}~'~\overline \xi_{1}=0, \quad
x_1~''-\exp(x_1- x_{0})-\xi_1~ \overline \xi_1~'=0;
\end{eqnarray}
\begin{eqnarray}
\xi_0=\overline \xi_2=\exp(-x_{-1})=\exp(x_2)=0;
\label{endb}
\end{eqnarray}
\begin{eqnarray}
x_0=x_0^{(0)}+x_0^{(2)}+x_0^{(4)}, \quad
x_1=x_1^{(0)}+x_1^{(2)}+x_1^{(4)}, \quad
\xi_1=\xi_1^{(1)}+\xi_1^{(3)}, \quad
\overline \xi_1=\overline \xi_1^{(1)}+\overline \xi_1^{(3)}, ~ ~ ~ ~ ~
\end{eqnarray}
respectively.

According to formulae (\ref{sol}), the zero-order
functions $x_0^{(0)}$ and $x_1^{(0)}$ are given by
\begin{eqnarray}
\exp (x_0^{(0)})=a_1\exp (b_1z)+a_2\exp (b_2z), \quad \exp(x_1^{(0)})=
-a_1a_2(b_1-b_2)^2\exp (z(b_1+b_2)-x^{(0)}_0). ~ ~ ~ ~ ~ ~ ~ ~ ~
\end{eqnarray}
Substituting them into eqs. (\ref{mm1})--(\ref{mm3}), taking into account
eqs. (\ref{endb}) and integrating the obtained expressions, we obtain the
following results:
\begin{eqnarray}
\xi_1^{(1)}={\beta}_1 x_1^{(0)}~'+{\alpha}_1 \exp (-x_0^{(0)}), \quad
\overline \xi_1^{(1)}=\overline {\beta}_1 x_0^{(0)}~'+
\overline {\alpha}_1 \exp(x_1^{(0)})
\end{eqnarray}
for the first-order functions $\xi_1^{(1)}$ and $\overline \xi_1^{(1)}$.

According to formulae (\ref{monoms}), we have
\begin{eqnarray}
&& X^{(2)}_0=\xi_1^{(1)}~'~\overline \xi_1^{(1)}, \quad
X^{(2)}_1=\xi_1^{(1)}~\overline \xi_1^{(1)}~', \nonumber\\
&& \Xi^{(3)}_{1}=-\xi_1^{(1)}~'~x_0^{(2)}~'+\xi_1^{(1)}(x_1^{(2)}-x_0^{(2)})
\exp(x_1^{(0)}- x_{0}^{(0)}), \nonumber\\
&& \overline \Xi^{(3)}_1=\overline \xi_1^{(1)}~'~x_1^{(2)}~'+
\overline \xi_1^{(1)}(x_1^{(2)}-x_0^{(2)})
\exp(x_1^{(0)}- x_{0}^{(0)}), \nonumber\\
&& X^{(4)}_0=-\frac{1}{2}(x_1^{(2)}-x_0^{(2)})^2\exp(x_1^{(0)}-x_{0}^{(0)})+
\xi_1^{(1)}~'~\overline \xi_1^{(3)}+\xi_1^{(3)}~'~\overline \xi_1^{(1)},
\nonumber\\
&& X^{(4)}_1=\frac{1}{2}(x_1^{(2)}-x_0^{(2)})^2\exp(x_1^{(0)}-x_{0}^{(0)})+
\xi_1^{(1)}~\overline \xi_1^{(3)}~'+\xi_1^{(3)}~\overline \xi_1^{(1)}~'
\label{monomse}
\end{eqnarray}
for the inhomogeneous parts of perturbative equations
(\ref{2lord})--(\ref{2l1ord2}), corresponding to the higher order
corrections. Iterating eqs. (\ref{n4}) and (\ref{intbbb}) in
consecutive order, step by step, we obtain the following expressions:
\begin{eqnarray}
&& x_0^{(2)}=-{1\over 2}({\beta}_1 \overline {\beta}_1-{\alpha}_1
\overline {\alpha}_1) x_1^{(0)}~' -
{\alpha}_1 \overline {\beta}_1 \exp (-x_0^{(0)}), \nonumber\\
&& x_1^{(2)}={1\over 2}({\beta}_1 \overline {\beta}_1-
{\alpha}_1 \overline {\alpha}_1)x_0^{(0)}~'+
{\beta}_1 \overline {\alpha}_1 \exp (x_1^{(0)})
\end{eqnarray}
for the second-order functions,
\begin{eqnarray}
&& \xi_1^{(3)}=-{1\over 2} {\alpha}_1 {\beta}_1
\overline {\beta}_1 (\exp (-x_0^{(0)}))~'+
{\alpha}_1 {\beta}_1 \overline {\alpha}_1
({1\over 2}\exp (x_1^{(0)}-x_0^{(0)})-b_1b_2), \nonumber\\
&& \overline \xi
_1^{(3)}=-{\alpha}_1 \overline {\alpha}_1~
\overline {\beta}_1 ({1 \over 2}
\exp (x_1^{(0)}-x_0^{(0)})-b_1 b_2)-
{1\over 2}{\beta}_1 \overline {\alpha}_1 ~
\overline {\beta}_1 (\exp (x_1^{(0)}))~'
\end{eqnarray}
for the third order functions, and, at last,
\begin{eqnarray}
&& x_0^{(4)}=-x_1^{(4)}=
{1\over 2} {\alpha}_1 {\beta}_1 \overline {\alpha}_1~\overline {\beta}_1
({1\over 2}\exp (x_1^{(0)}-x_0^{(0)})- b_1 b_2)
\end{eqnarray}
for the fourth order functions.

\section{Conclusion}

In the present Letter, we proved the integrability  of the f-Toda chain
with fixed ends or interrupted from the left (right) and proposed an
algorithmic method for constructing its explicit solutions. Many
interesting questions arise in this connection.  What is the
group-theoretical foundation of this result and its connection with the
representation theory of supergroups (superalgebras)? Is it possible to
construct a superintegrable two-dimensional generalization of the f-Toda
chain? Is it possible to represent the f-Toda chain with fixed ends in the
Lax-pair or Hamiltonian forms? This last question is a very important in
connection with the problem of its quantization. In the case of the usual
Toda chain, all these questions have the answers and the authors hope to
find their solutions for the case of the f-Toda chain in future
publications.

\section*{Acknowledgments}

One of us (A.S.) would like to thank E. Ivanov for his interest in this
work and useful discussions. This work was partially supported by the
Russian Foundation for Basic Research, Grant 96-02-17634, INTAS Grant
94-2317, and by a grant from the Dutch NWO organization.

\end{document}